\documentclass[a4paper,USenglish,cleveref,autoref,thm-restate]{lipics-v2021}

\usepackage{booktabs}
\usepackage{siunitx}
\usepackage{graphicx}
\usepackage{xcolor}
\usepackage{framed}
\definecolor{shadecolor}{gray}{0.95}
\newenvironment{rqbox}{
  \par\noindent
  \setlength{\FrameRule}{0.4pt}
  \setlength{\FrameSep}{6pt}
  \begin{shaded*}\noindent
}{
  \end{shaded*}
}
\newcommand{\dataset}{\textsc{CodAGE}}
\title{AI-to-AI Code Reviews of GitHub Pull Requests}

\author{Niruthiha Selvanayagam}{\'Ecole de technologie sup\'erieure (\'ETS Montr\'eal), Montr\'eal, Canada}{niruthiha.selvanayagam.1@ens.etsmtl.ca}{0000-0002-7853-3048}{}
\author{Taher A. Ghaleb}{Department of Computer Science, Trent University, Peterborough, Canada}{taherghaleb@trentu.ca}{0000-0001-9336-7298}{}

\authorrunning{N. Selvanayagam and T.\,A. Ghaleb}
\Copyright{Niruthiha Selvanayagam and Taher A. Ghaleb}

\funding{This work is partially supported by the Natural Sciences and Engineering Research
Council of Canada (NSERC): RGPIN-2025-05897.}

\EventEditors{Robert Feldt, Maria Paasivaara, Daniel Mendez, Stefan Wagner, and Marvin Mu\~{n}oz Bar\'{o}n}
\EventNoEds{5}
\EventLongTitle{20th International Symposium on Empirical Software Engineering and Measurement (ESEM 2026)}
\EventShortTitle{ESEM 2026}
\EventAcronym{ESEM}
\EventYear{2026}
\EventDate{October 8--9, 2026}
\EventLocation{Munich, Germany}
\EventLogo{}
\SeriesVolume{394}
\ArticleNo{74}

\category{Emerging Results, Vision \& Reflection Track Paper}



\supplementdetails{(Replication Package)}{https://github.com/Niruthiha/AI-AI-CodeReviews}

\nolinenumbers

\begin{document}

\maketitle

\begin{abstract}
AI coding agents are increasingly integrated into software development workflows, operating on both sides of the pull request (PR) process: \emph{AI authoring agents}, which create or modify PRs, and \emph{AI reviewers}, which evaluate them. This creates a closed loop where one AI coding agent reviews contributions of another AI coding agent.
In this paper, we construct an AI-to-AI code review dataset by linking AI-authored pull requests with AI-attributed review events from \dataset, a public dataset of coding agent–generated GitHub events. Our dataset contains \num{248641} unique AI-attributed PRs that received at 
least one AI-attributed review. Among these, \num{45269} received cross-product 
review and \num{208145} received same-product review; \num{4773} PRs received both.
We observe that cross-product AI-to-AI code review occurs in only about 1.6\% of identified agent-authored PRs but is substantial in absolute terms: 45k PRs written by one identifiable AI product and reviewed by another. This activity grows by more than two orders of magnitude from 2025-Q1 to 2025-Q3.
We measure reviewer behavior using CodeRabbit comment categories, per-PR comment volume, and time to first review, and find that it varies across author–reviewer pairs. For example, Claude-Code PRs receive more \texttt{refactor} comments from CodeRabbit than Copilot PRs (35.0\% vs.\ 10.5\%), a difference that may stem from PRs themselves rather than the reviewer. For three of four dual-role reviewers, mean comments per PR were 58--65\% higher in the same-product group, though effect sizes were small or negligible and the difference was concentrated in the upper tail. Among PR--reviewer pairs with complete, nonnegative timestamps, median time 
from PR creation to first AI review was \num{1.2} minutes for cross-product 
pairs and \num{4.7} minutes for same-product pairs. This descriptive difference 
may reflect reviewer composition and differential timestamp availability rather 
than product pairing itself.
Overall, our large-scale characterization shows that closed-loop AI-to-AI code review is on the rise but remains a minority phenomenon, with review output varying across authoring-agent groups and author--reviewer configurations.
\end{abstract}

\section{Introduction}
\label{sec:intro}

Two types of AI coding agents now commonly appear in GitHub pull-request workflows. Some agents author code: systems such as Devin, OpenAI Codex, Copilot, Cursor, and Claude Code can open pull requests (PRs) on behalf of users, either autonomously or as assistive tools~\cite{li2026aidev}. Others review code: tools such as CodeRabbit, Sourcery, and PR-Agent post line-level comments and review decisions. In addition, several authoring agents, including Copilot, Codex, Gemini Code Assist, and Amazon Q, also appear as reviewers in public GitHub data.

Together, these roles form what we call \emph{closed-loop AI review}: an AI coding agent contributes to a GitHub repository, and one or more AI coding agents review it. We use ``closed-loop'' in this observable sense only. Our review stream is restricted to AI-attributed review events, so the term means that AI occupies both sides of the pull request, not that humans were absent: a PR in our datasets may also have received human review that we do not observe. 
This shift raises two research questions (RQs):

\begin{itemize}
    \item \textbf{RQ1: How prevalent is closed-loop AI review on GitHub?}
    
    This RQ characterizes the prevalence and growth of closed-loop review. As AI-generated reviews become more common, an increasing share of public code may be evaluated partly by AI, affecting both downstream maintenance and studies that rely on GitHub review data as evidence of human decision making.
    
    \medskip
    
    \item \textbf{RQ2: How does reviewer behavior vary with the author--reviewer setup?}
    
    This RQ examines whether the authoring agent influences reviewer behavior. If reviewer behavior varies across authoring agents, then closed-loop review cannot be treated as a single process. Instead, both maintainers and researchers need to account for which agent wrote the code and which agent reviewed it.

\end{itemize}

Prior work has primarily studied AI coding tools and AI review systems in isolation, including productivity effects of code generation tools~\cite{peng2023copilot} and usability of AI-assisted suggestions~\cite{vaithilingam2022expectation}. Earlier studies of reviewer bots~\cite{wessel2020effects} were conducted before the current ecosystem of dual-role AI coding agents and do not analyze author--reviewer pairings.
This paper addresses these questions by introducing a large-scale dataset and empirical study of AI-to-AI code reviews of GitHub PRs. We identify AI-authored and AI-reviewed PRs by applying our attribution method to \dataset~(CodAGE) events~\cite{codage} from 2024--2026, enabling measurement of closed-loop interactions at scale.
We find this phenomenon already substantial in absolute terms, if still a minority of all agent activity: in 2025, tens of thousands of PRs authored by AI coding agents are reviewed by other AI coding agents across multiple products.

\medskip\noindent\textbf{Contributions.}
Overall, this paper makes the following contributions.

\begin{enumerate}
    \item We introduce a large-scale dataset of AI-to-AI code reviews, covering \num{248641} PRs that received at least one AI review: \num{45269} cross-product and \num{208145} same-product.
    
    \item We characterize closed-loop AI code review patterns over time, tracking how PRs are reviewed by the same or different AI product.
    
    \item We empirically analyze AI reviewer behavior under different author--reviewer setups, examining how reviewer output varies with authoring agent and setup using CodeRabbit comment-category labels, per-pull-request comment volume, and time to first review.
\end{enumerate}

\medskip\noindent\textbf{Paper Organization.}
The rest of this paper is organized as follows. Section~\ref{sec:background} reviews related work. Section~\ref{sec:data} describes our data and methodology. Sections~\ref{sec:rq1} and~\ref{sec:rq2} present our findings on closed-loop AI review prevalence and reviewer behavior. Section~\ref{sec:discussion} discusses implications.
Section~\ref{sec:threats} addresses threats to validity.
Section~\ref{sec:future} concludes the paper and outlines future directions.

\section{Background and Related Work}
\label{sec:background}

\smallskip\noindent\textbf{AI-authored pull requests on GitHub.}
Prior work on AI code generation has largely focused on code-suggestion tools used by human developers, including studies of Copilot productivity~\cite{peng2023copilot}, usability of LLM-based suggestions~\cite{vaithilingam2022expectation}, and AI-assisted pair programming~\cite{imai2022copilot}. More recent autonomous or semi-autonomous agents open PRs with limited or no line-by-line human authorship~\cite{li2026aidev}. Prior work characterizes these agent-authored PRs directly: in one study of Claude Code, \SI{45.1}{\percent} of merged PRs required additional changes~\cite{watanabe2025agentic}; non-merged agent-authored PRs tend to involve larger changes, touch more files, and fail CI more often~\cite{ehsani2026wheredo}; and merge outcomes turn on reviewer intervention and repository context more than on agent identity alone~\cite{selvanayagam2026agent}. This work centers on whether agent-authored code is accepted; it says little about how AI reviewers behave when they are the ones doing the evaluating.

\smallskip\noindent\textbf{Reviewer-side bots.}
Wessel et al.~\cite{wessel2020effects} characterized the adoption of bots on GitHub and showed that bot introduction can affect human review volume and latency. Brown and Parnin~\cite{brown2019sorry} studied developer interaction with bots in code review workflows, and recent surveys chart how generative AI is now reshaping modern code review more broadly~\cite{yang2026roadmap}. LLM-based review has since moved into practice, with systems evaluated or deployed at industrial scale~\cite{li2022codereviewer,cihan2025acrpractice,sun2025bitsaicr,tantithamthavorn2026rovodev} and multi-agent review pipelines developed in research~\cite{tang2024codeagent,ren2025hydra}, alongside a parallel line of work asking whether their output is trustworthy: developers accept only a minority of CodeRabbit suggestions~\cite{lin2026coderabbit}, current systems remain weak at realistic review generation~\cite{zeng2026swrbench,zhou2023generationbased}, and review comments can be ungrounded in the code they describe~\cite{tantithamthavorn2026hallujudge}. A closer group of recent studies looks at reviewer bots specifically on agent-authored PRs~\cite{zhong2026synergy,zhong2026humanagentic}, reporting that AI suggestions are adopted far less often than human ones and that agent involvement can speed decisions without reliably improving review quality. This work establishes that AI now reviews AI-authored code; what remains open, and what we take up, is which agents review which others across the ecosystem, and whether observable reviewer behavior differs between same-product and cross-product configurations.

\smallskip\noindent\textbf{Bot/Agent attribution.}
Prior work identifies automated accounts using commit- and account-level heuristics, such as BIMAN~\cite{dey2020biman}, or comment-level classifiers, such as BoDeGHa~\cite{golzadeh2021bodega}. Ghaleb et al.~\cite{ghaleb2026fingerprinting} recently used machine learning models to detect AI coding agents using behavioral fingerprints. While these studies distinguish automated contributions or infer the originating agent, they do not identify the corresponding AI product or vendor, a gap we address using signature-based attribution.

\smallskip\noindent\textbf{Multi-agent LLM evaluation and this work.}
Outside software engineering, ``LLM-as-a-judge''~\cite{zheng2023llmjudge} and constitutional AI~\cite{bai2022constitutional} study settings in which models evaluate, critique, or revise model-generated outputs. However, the corresponding deployed software setting, AI reviewers commenting on AI-authored PRs in public repositories, is far less studied. To our knowledge, prior work has not characterized product-resolved AI-author--AI-reviewer pairings at this scale across GitHub. We address this gap by quantifying how often such closed-loop interactions occur, which agents participate, and how reviewer behavior varies with the authoring agent and the author--reviewer configuration.

\section{Data and Methodology}
\label{sec:data}

This section describes the raw GitHub events data, the attribution of events to AI coding agents, the analysis dataset, and the CodeRabbit comment-category classifier.

\subsection{Data source}
\label{sec:data:source}

Our data source is the \dataset~dataset (Coding Agent-generated GitHub Events)~\cite{codage} collected from GHArchive.\footnote{\url{https://www.gharchive.org}} We use the snapshot covering events from 2024-01-01 to 2026-04-15. \dataset~contains several event-type subsets. In this paper, we focus on CodAGE-PRs, CodAGE-Reviews, and CodAGE-ReviewComments from this period. These subsets capture PR authorship signatures and reviewer-side activity. Despite covering the relevant period, many autonomous AI coding agents emerged in early 2025. Candidate events were identified by a broad attribution pass over CodAGE-PRs, which includes branch-name evidence and yields \num{4563819} candidate PRs with a provisional agent label. The signature framework described in Section~\ref{sec:data:signatures} is a stricter second pass over that pool: it requires body or vendor-login evidence and discards branch-only matches.

\subsection{Identifying AI authorship and AI review}
\label{sec:data:signatures}

Attributing PRs to AI coding agents by actor login alone is unreliable: some use generic \texttt{[bot]} accounts, some use the human user's login with a machine-readable body signature, and some quote other agents' text, which can mislead body-only rules. We therefore use a two-tier signature framework. \textbf{S1 (high-confidence body signature)} is a regex or fixed substring the agent produces in the PR or comment body, such as the \texttt{Co-Authored-By: Claude <noreply@anthropic.com>} trailer from Claude Code, the \texttt{cursor.com/agents} URL from Cursor, or the \texttt{chatgpt.com/codex} URL from OpenAI Codex. \textbf{S2 (vendor-controlled account signature)} is an exact match or regex over a login unambiguously controlled by a vendor, such as \texttt{coderabbitai[bot]}, \texttt{devin-ai-integration[bot]}, or \texttt{gemini-code-assist[bot]}. \texttt{S1} agents require body-level evidence; \texttt{S2} agents accept body or login evidence. Branch-name prefixes (e.g.\ \texttt{codex/}, \texttt{cursor/}) are recorded but never sufficient alone, being user-controllable and prone to false positives.

We create one entry per matched (event, agent) pair. A few events match multiple agents, usually when one agent quotes another's body trailer: 98 of \num{4141107} review entries (\SI{0.002}{\percent}) and 111 of \num{8560237} review-comment entries (\SI{0.001}{\percent}). We retain all matches, report the duplicate rate in the coverage report, and deduplicate by event ID when event-grain analysis requires it. This attributes \emph{review-side} events to 12 AI coding agents (CodeRabbit, Copilot, Gemini Code Assist, OpenAI Codex, Amazon Q, Devin, Claude Code, Sweep AI, PR-Agent, Kiro, Cursor, Aider). The \emph{author-side} stream yields \num{2830284} unique agent-authored PRs and adds Google Jules, which authors PRs but has no reviewer activity.

\subsection{Cleaning, deduplication, and quarantine}
\label{sec:data:clean}

We apply three cleaning steps.

\emph{\textbf{(i) Quarantine.}} Every candidate PR or review event is scored against each agent's signature set, and an entry is emitted only where the evidence meets that agent's \emph{inclusion floor}: agents with an S1 floor require a body signature, whereas agents with an S2 floor accept either a body signature or a vendor-controlled login. A vendor-controlled login is therefore sufficient on its own for S2-floor agents. Any candidate meeting no agent's floor is quarantined, written to a separate file with a recorded reason. On the author side the reasons are \texttt{branch\_only}, where only a branch-name prefix matched, and \texttt{no\_evidence}; on the review side, they are \texttt{empty\_body} and \texttt{no\_evidence}. Quarantine is negligible on the two review streams (\num{590} of \num{4141598} review events and \num{793} of \num{8560919} review comments, \SI{0.01}{\percent} of each) but substantial on the author stream, where \num{1733535} of \num{4563819} candidate PRs (\SI{38.0}{\percent}) are quarantined, \SI{96.9}{\percent} of them as \texttt{branch\_only}. This asymmetry underlies the coverage imbalance discussed in Section~\ref{sec:threats}: our agent-authored population is the \SI{62.0}{\percent} of candidates with a body or login signature, not all PRs a weaker rule would include.

\emph{\textbf{(ii) Relabeling.}} We retain the provisional label from the first pass as \texttt{original\_label} and report the two passes as a confusion matrix. Where they disagree, the signature attribution wins; the largest relabel consolidates \texttt{OpenAI\_Codex\_Cloud} and the CLI Codex trailer into \texttt{OpenAI\_Codex}. Because both passes are ours, this is an internal consistency check rather than external validation.

\emph{\textbf{(iii) Thin agents.}} Agents with fewer than 50 attributed PRs (Aider, Kiro, SWE-agent, Windsurf) are excluded from our per-agent quantitative claims.

\subsection{Dataset construction}
\label{sec:data:construction}

The unit of analysis is the \emph{PR}, keyed by \texttt{(repo\_name, pr\_number)}. We aggregate per-event entries into one entry per (PR, Reviewer) pair, retaining review and review-comment counts and first/last review timestamps, then join with the agent-author PR set to yield three datasets (\texttt{A}: AI author; \texttt{B}: AI reviewer; \texttt{N}: no \emph{detected} AI reviewer): \textbf{A$\times$B-cross-product}, the closed-loop cross-product dataset with author and reviewer from different products (\num{45269} PRs, \num{47259} pairs); \textbf{A$\times$B-same-product}, same product, e.g.\ Copilot reviewing Copilot (\num{208145} PRs); and \textbf{A$\times$N}, agent-authored PRs with no \emph{detected} AI review (\num{2581643} PRs, for context only); absence of an attributed reviewer signature is not evidence that no AI reviewed the PR. A$\times$B PRs receiving both same- and cross-product review appear in both A$\times$B datasets with an overlap flag, so the dataset counts are not mutually exclusive and do not sum to the unique agent-authored total. We do \emph{not} analyze a matched human-authored control dataset.

These two closed-loop datasets are defined at the level of the identifiable agentic \emph{product}, or agentic harness, rather than the parent company or the underlying foundation model: \textbf{A$\times$B-same-product} contains PRs in which the same identifiable product appears as author and reviewer, and \textbf{A$\times$B-cross-product} those in which the authoring and reviewing products differ. Products operated by the same corporate vendor, such as Google Jules and Gemini Code Assist, are therefore treated as distinct, because they are separately identifiable agentic systems with different product names and potentially different harnesses, configurations, and review workflows; \num{117} of the \num{45269} A$\times$B-cross-product PRs are of this kind. These categories consequently do not necessarily correspond to corporate vendor or foundation-model boundaries.

\subsection{Severity classifier and released artifacts}
\label{sec:data:severity}

\textbf{CodeRabbit comment categories.} To measure \emph{review content} for RQ2, we use CodeRabbit's self-declared comment categories. CodeRabbit prefixes substantive comments with emoji-tagged headers: \textbf{Refactor suggestion}, \textbf{Potential issue}, and \textbf{Nitpick}, plus a fourth header, \textbf{Verification}, which appears in our historical data (\num{9525} comments) and which we treat as an observed comment prefix rather than a documented category. Because these headers appear explicitly in the comment text, we extract them deterministically using rule-based matching without an LLM judge. We retain \emph{severity} in artifact and column names for consistency with the released dataset, but do not interpret the categories as an ordered severity scale. In our dataset, CodeRabbit is the only dedicated reviewer-only bot with both substantial volume and machine-parsable category headers (Sourcery and PR-Agent together account for fewer than 250 review events).

\medskip\noindent\textbf{Classifier.} An ordered-regex pipeline strips boilerplate that would otherwise produce false matches, then applies the rules in order to the cleaned comment body; the first rule wins, and unmatched comments are \texttt{unlabeled}. It emits four substantive categories, \texttt{potential\_issue} (suspected bug), \texttt{refactor} (working code whose structure could improve), \texttt{verification} (a request to verify a behavior or assumption), and \texttt{nitpick} (minor style), on which we impose no ordering, and three structural labels, \texttt{analysis\_chain} (script-execution diagnostic), \texttt{conversational} (\texttt{@}-mention reply), and \texttt{rate\_limited} (an operational rate-limit notice). The rule set and its ordering are provided in the replication package.

\medskip\noindent\textbf{Replication package.} Our replication package (data processing/analysis scripts, filtered data, signature registry, and detailed results) is available on GitHub.\footnote{\url{https://github.com/Niruthiha/AI-AI-CodeReviews}} 
The main source data used in this work is \dataset~(Coding Agent-generated GitHub Events), hosted on Hugging Face,\footnote{\url{https://huggingface.co/datasets/taher-ghaleb/CodAGE}} specifically CodAGE-PRs, CodAGE-Reviews, and CodAGE-ReviewComments.

\section{RQ1: Prevalence and Composition of Closed-Loop AI-to-AI Code Reviews}
\label{sec:rq1}

We first establish the scale and growth of closed-loop review, then decompose it by author--reviewer pair and by same- versus cross-product composition.

\subsection{Volume and growth}
\label{sec:rq1:volume}

\begin{figure}[ht]
    \centering
    \includegraphics[width=0.85\linewidth]{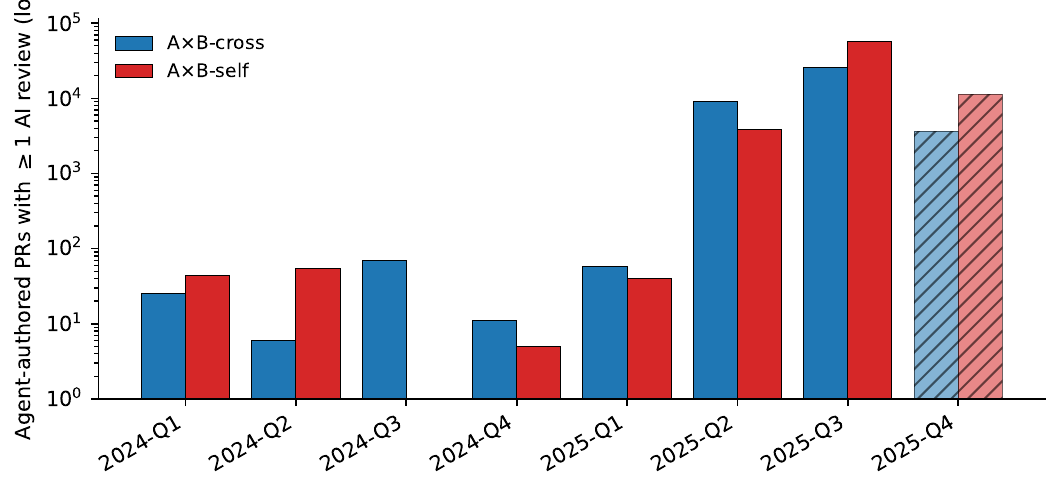}
    \caption{Quarterly counts of agent-authored PRs receiving AI review, split into cross-product (A$\times$B-cross-product) and same-product (A$\times$B-same-product) groups. Both increased by over two orders of magnitude from 2025-Q1 to 2025-Q3. Hatched 2025-Q4 values are lower bounds because GHArchive attributes lag event data by about one quarter. The y-axis is log-scaled.}
    \label{fig:rq1-time}
    \vspace{-5pt}
\end{figure}

Of the \num{2830284} agent-authored PRs we identified, \num{248641} (\SI{8.8}{\percent}) received at least one AI review: \num{45269} cross-product and \num{208145} same-product, with the counts read as lower bounds and the percentage as conditioned on signature attribution (Section~\ref{sec:threats}). We focus on cross-product review because same-product review may be part of an integrated product workflow. Activity was negligible through 2025-Q1 (\num{57} cross, \num{40} same) but increased by over two orders of magnitude by 2025-Q3, reaching \num{25492} cross- and \num{57080} same-product PRs (Figure~\ref{fig:rq1-time}). Because 2025-Q4 counts are incomplete due to GHArchive attribute lag, temporal comparisons use Q2 and Q3.

\subsection{Who reviews whom?}
\label{sec:rq1:matrix}

Table~\ref{tab:rq1-matrix} reports the A$\times$B-cross-product author--reviewer crosstab ($N \geq 500$). OpenAI Codex dominates as author (\num{31601} cross-product PRs, \SI{69.8}{\percent} of the \num{45269}). Its row in Table~\ref{tab:rq1-matrix} sums to \num{32379} rather than \num{31601} because those cells count author--reviewer pairs, not unique PRs: a PR reviewed by agents from two different vendors counts once in Table~\ref{tab:rq1-vendor} but contributes a pair to each reviewer column here (and the crosstab shows only cells with $N \geq 500$). Copilot dominates as reviewer (\num{21022} of the \num{47259} author--reviewer pairs; modal pair \emph{Codex authored, Copilot reviewed}, \num{18114} pairs), and CodeRabbit, the only dedicated reviewer-only bot with usable volume, reviews PRs from at least six authoring agents.

\begin{table}[ht]
    \centering
    \caption{Closed-loop, cross-product author--reviewer pairs (A$\times$B-cross-product), cells with $N \geq 500$. Cells report author--reviewer pairs. Because one PR may receive reviews from multiple agents, row totals may exceed the corresponding number of unique PRs.}
    \label{tab:rq1-matrix}
    \small
    \begin{tabular}{l r r r r r}
        \toprule
        \textbf{Author $\rightarrow$ Reviewer} & \textbf{Copilot} & \textbf{Codex} & \textbf{Gemini} & \textbf{CodeRabbit} & \textbf{Total ($\geq$ 500)} \\
        \midrule
        OpenAI Codex      & \num{18114} &       --     & \num{8995} & \num{5270} & \num{32379} \\
        Copilot           &       --    & \num{4772} & \num{956}  & \num{1627} & \num{7355} \\
        Cursor            & \num{1000}  &       --     & \num{593}  & \num{516}  & \num{2109} \\
        Claude Code       & \num{956}   &       --     & \num{501}  & \num{716}  & \num{2173} \\
        Devin             & \num{656}   &       --     &      --    &      --    & \num{656} \\
        \bottomrule
    \end{tabular}
\end{table}

\begin{table}[ht]
    \centering
    \caption{Same- vs.\ cross-product reviewer share, per authoring agent (A$\times$B-cross-product + A$\times$B-same-product, $\geq 200$ PRs). Sorted by total A$\times$B in descending order.}
    \label{tab:rq1-vendor}
    \small
    \setlength{\tabcolsep}{2pt}
    \begin{tabular}{l r r r r}
        \toprule
        \textbf{Authoring agent} & \textbf{Cross-product PRs} & \textbf{Same-product PRs} & \textbf{Total A$\times$B} & \textbf{\% Same-product} \\
        \midrule
        Copilot            & \num{7515}  & \num{166442} & \num{173957} & \SI{95.7}{\percent} \\
        OpenAI Codex       & \num{31601} & \num{37247}  & \num{68848}  & \SI{54.1}{\percent} \\
        Devin              & \num{1171}  & \num{3811}   & \num{4982}   & \SI{76.5}{\percent} \\
        Cursor             & \num{2309}  & 0            & \num{2309}   & \SI{0.0}{\percent} \\
        Claude Code        & \num{2042}  & \num{10}     & \num{2052}   & \SI{0.5}{\percent} \\
        Amazon Q           & \num{48}    & \num{535}    & \num{583}    & \SI{91.8}{\percent} \\
        Google Jules       & \num{478}   & 0            & \num{478}    & \SI{0.0}{\percent} \\
        Sweep AI           & \num{100}   & \num{100}    & \num{200}    & \SI{50.0}{\percent} \\
        \bottomrule
    \end{tabular}
\end{table}

\subsection{Same- vs.\ cross-product composition}
\label{sec:rq1:vendor}

Table~\ref{tab:rq1-vendor} reports the same- vs.\ cross-product split per authoring agent ($\geq 200$ A$\times$B PRs). The pattern is sharply non-uniform, with three groups: \emph{product-internal} authors reviewed mostly by their own product (Copilot \SI{95.7}{\percent}, Amazon Q \SI{91.8}{\percent}, Devin \SI{76.5}{\percent} same-product); \emph{balanced} (OpenAI Codex at \SI{54.1}{\percent} self, the only author with substantial population in both datasets); and \emph{cross-product} authors reviewed almost entirely by a different product (Cursor, Google Jules, Claude Code). This asymmetry motivates RQ2, in which we (i) hold the reviewer constant (CodeRabbit) and vary the author, (ii) contrast same- vs.\ cross-product review volume, and (iii) contrast time-to-first-review across reviewer bots and product pairings.

\vspace{-10pt}
\begin{rqbox}
\textbf{RQ1 Summary:} AI review remains a minority of agent activity but is already substantial: \num{248641} of the agent-authored PRs we identified (\SI{8.8}{\percent}) received at least one AI review, including \num{45269} (\SI{1.6}{\percent}) across products, spanning \num{10345} repositories. The counts are lower bounds; the percentages are conditioned on signature attribution. Activity grew by over two orders of magnitude in 2025. Copilot-authored PRs were reviewed mainly by Copilot, whereas Cursor-, Claude Code-, and Google Jules-authored PRs were reviewed almost entirely across products.
\end{rqbox}

\section{RQ2: Reviewer Behavior Across Authors and Product Pairings}
\label{sec:rq2}

We characterize reviewer behavior using three measures. First, we examine CodeRabbit's \emph{comment-category mix} while holding the reviewer fixed and varying the authoring agent. Because these are self-declared labels rather than validated defects, we interpret them as reviewer output composition, not code quality~\cite{lin2026coderabbit,tantithamthavorn2026hallujudge,jin2026reliable}. Second, we compare per-PR comment \emph{volume} between same- and cross-product pairs for Copilot, OpenAI Codex, Devin, and Amazon Q. Third, we descriptively examine latency from PR creation to the first AI review 
among PR--reviewer pairs with complete, nonnegative timestamps, considering 
whether aggregate differences may reflect reviewer composition or differential 
timestamp availability rather than product pairing.

\subsection{CodeRabbit comment categories by authoring agent}
\label{sec:rq2:contentcut}

\begin{figure}[ht]
    \centering
    \includegraphics[width=0.95\linewidth]{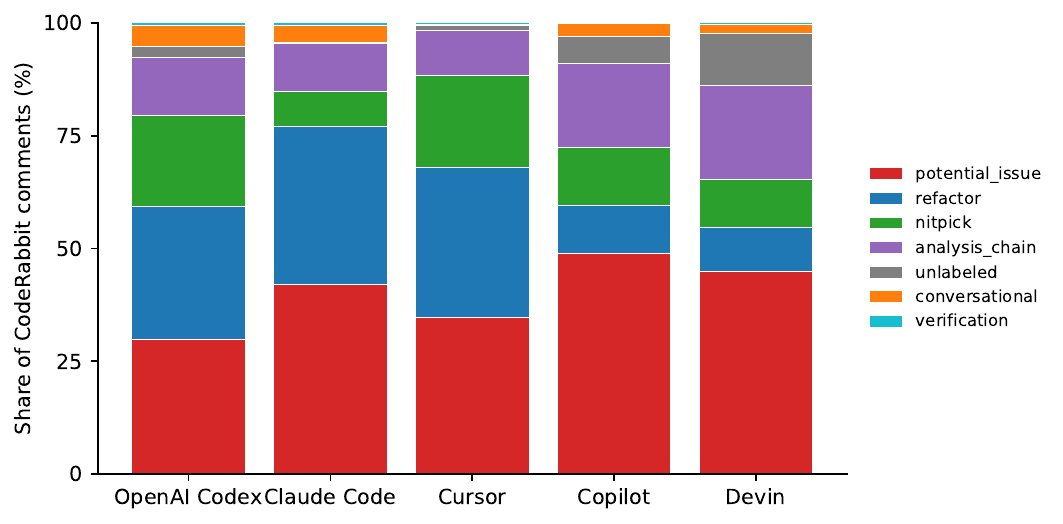}
    \caption{Normalized distribution of CodeRabbit comment categories on agent-authored PRs, grouped by authoring agent (bars sum to 100\%). The same reviewer bot produces visibly different mixes downstream of different authoring agents; the largest single difference is \texttt{refactor}, contrasting the Claude Code / Cursor / Codex cluster (29--35\%) with the Copilot / Devin pair (10--11\%).}
    \label{fig:rq2-severity}
    \vspace{-5pt}
\end{figure}

We join the CodeRabbit classifier output against the union of the A$\times$B-cross-product and A$\times$B-same-product datasets on \texttt{(repo\_name, pr\_number)}, restricting to the five authoring agents with $\geq 200$ classified comments (\num{35248} comments total; Figure~\ref{fig:rq2-severity} and Table~\ref{tab:rq2-cutA}). The comment-category distribution is visibly non-uniform: Copilot- and Devin-authored PRs receive far more \texttt{potential\_issue} comments (\SI{49.0}{\percent}, \SI{43.9}{\percent}) and far fewer \texttt{refactor} comments (\SI{10.5}{\percent}, \SI{9.7}{\percent}) than Claude Code, Cursor, and OpenAI Codex (\SIrange{29.2}{35.0}{\percent} \texttt{refactor}), and the highest \texttt{unlabeled} shares (\SI{6.1}{\percent}, \SI{12.3}{\percent}). Claude Code sits at the opposite end, with the highest \texttt{refactor} (\SI{35.0}{\percent}) and lowest \texttt{unlabeled} (\SI{0.3}{\percent}).

\begin{table}[ht]
    \centering
    \caption{Normalized distribution of CodeRabbit comment categories (\%) on agent-authored PRs, authors with $N \geq 200$ classified comments. Last column is the total number of entries ($N$).}
    \label{tab:rq2-cutA}
    \small
    \begin{tabular}{l r r r r r r r r}
        \toprule
        \textbf{Author} & \textbf{p.issue} & \textbf{refactor} & \textbf{nitpick} & \textbf{a.chain} & \textbf{unlab.} & \textbf{conv.} & \textbf{verif.} & \textbf{N} \\
        \midrule
        OpenAI Codex & 29.9 & 29.2 & 20.5 & 12.8 & 2.4  & 4.7 & 0.5 & \num{16292} \\
        Claude Code  & 42.0 & 35.0 &  7.8 & 10.6 & 0.3  & 3.7 & 0.6 & \num{8757} \\
        Cursor       & 34.9 & 33.5 & 19.8 & 10.2 & 0.9  & 0.2 & 0.5 & \num{5084} \\
        Copilot      & 49.0 & 10.5 & 12.9 & 18.9 & 6.1  & 2.5 & 0.1 & \num{4260} \\
        Devin        & 43.9 &  9.7 & 11.6 & 20.0 & 12.3 & 2.2 & 0.4 & \num{855} \\
        \bottomrule
    \end{tabular}
\end{table}

The association between authoring agent and CodeRabbit comment category is statistically significant but small (Cram\'er's $V = 0.150$; Pearson $\chi^2 = 3177.5$, $df = 24$, $p < 10^{-300}$). We therefore focus on the largest category-level contrast. CodeRabbit labels \SI{35.0}{\percent} of comments on Claude Code-authored PRs as \texttt{refactor}, compared with \SI{10.5}{\percent} on Copilot-authored PRs, a difference of 24.5 percentage points (95\% CI $[23.1, 25.9]$; Wald two-proportion test; $N = 8757$ and $N = 4260$ comments, respectively). We make no claim about code quality, as change size, language, repository, and configuration may influence the patterns observed. Our result is therefore observational and indicates that CodeRabbit produces different comment-category distributions for Claude Code- and Copilot-authored PRs.

\subsection{Same- vs.\ cross-product review volume}
\label{sec:rq2:volumecut}

We restrict to reviewer bots with $\geq 200$ PRs in \emph{both} datasets (Copilot, OpenAI Codex, Devin, Amazon Q) and compare per-PR comment counts on same- vs.\ cross-product authored PRs (Table~\ref{tab:rq2-cutB}; we report means and use Mann--Whitney $U$ tests to compare the distributions.

\begin{table}[ht]
    \centering
    \caption{Mean review comments per PR by reviewer, split by same- and cross-product authors. Counts use the reviewer perspective: ``Self PRs'' match the same-product totals in Table~\ref{tab:rq1-vendor}, whereas ``Cross PRs'' do not. $p$ values are from two-sided Mann--Whitney $U$ tests; $|\delta|$ is Cliff's delta (self vs.\ cross; $<0.147$ negligible, $0.147$--$0.33$ small).}
    \label{tab:rq2-cutB}
    \small
    \begin{tabular}{l r r r r r r}
        \toprule
        \textbf{Reviewer bot} & \textbf{Cross PRs} & \textbf{Mean cross} & \textbf{Self PRs} & \textbf{Mean self} & \textbf{$p$} & \textbf{$|\delta|$} \\
        \midrule
        Copilot       & \num{21022} & 1.49 & \num{166442} & 2.35 & $<10^{-292}$ & 0.15 \\
        OpenAI Codex  & \num{5485}  & 0.91 & \num{37247}  & 0.89 & $2.0 \times 10^{-3}$ & 0.02 \\
        Devin         & \num{392}   & 1.09 & \num{3811}   & 1.80 & $3.0 \times 10^{-6}$ & 0.14 \\
        Amazon Q      & \num{522}   & 4.94 & \num{535}    & 8.08 & $2.2 \times 10^{-14}$ & 0.27 \\
        \bottomrule
    \end{tabular}
\end{table}

The per-PR distributions overlap heavily and Cliff's $\delta$ is small-to-negligible across all four bots (Devin and Copilot sit within $0.01$ of the $0.147$ boundary, so labels are indicative only). The mean-level gap is therefore not a population-wide shift but a long right tail of high-comment same-product PRs: three of four bots show same-product PRs receiving 58--65\% more comments on average (Copilot \SI{58}{\percent}, 1.49 to 2.35; Devin \SI{65}{\percent}, 1.09 to 1.80; Amazon Q \SI{64}{\percent}, 4.94 to 8.08), while OpenAI Codex is the lone exception (0.91 vs.\ 0.89, too small for a directional claim despite a sample-size-driven $p$). This gap is consistent across three otherwise dissimilar products, even though it is concentrated in the upper tail of the distribution. Because same- and cross-product reviews differ in several key characteristics, including change size, language, repository, and reviewer composition, the observed differences should not be interpreted causally.

\subsection{Latency from PR open to first AI review}
\label{sec:rq2:latencycut}

We calculate the elapsed time between PR creation and the bot’s first recorded review event for PR–reviewer pairs with the required timestamps. After dropping pairs missing either timestamp and the small share with negative differences (clock-skew or backfill), \num{103920} pairs remain; distributions are heavily right-skewed, so we report medians with IQRs (Table~\ref{tab:rq2-cutC}).

\begin{table}[ht]
    \centering
    \caption{\caption{Median latency from PR creation to first AI review, in minutes, among 
PR--reviewer pairs with complete, nonnegative timestamps. Top: same- vs.\
cross-product pairs aggregated. Bottom: by reviewer bot with the datasets 
pooled, including reviewer bots with $\geq 100$ qualifying pairs.}}
    \label{tab:rq2-cutC}
    \footnotesize
    \begin{tabular}{l r r r r}
        \toprule
        \textbf{Split} & \textbf{N} & \textbf{p25} & \textbf{median} & \textbf{p75} \\
        \midrule
        Cross-product pairs & \num{37420} & 0.4 & 1.2 & 4.4 \\
        Same-product pairs  & \num{66500} & 2.4 & 4.7 & 38.2 \\
        \midrule
        Gemini Code Assist (reviewer) & \num{9873}  & 0.3 &  0.5 &  1.8 \\
        Amazon Q (reviewer)           & \num{959}   & 0.9 &  1.2 &  1.6 \\
        OpenAI Codex (reviewer)       & \num{34765} & 1.9 &  2.6 &  3.7 \\
        CodeRabbit (reviewer)         & \num{6709}  & 3.7 &  7.0 & 31.6 \\
        Copilot (reviewer)            & \num{51437} & 1.0 & 17.7 & 73.9 \\
        \bottomrule
    \end{tabular}
\end{table}

Two observations. First, median first-review latency in the aggregate is 1--5 minutes; AI review on these PRs commonly begins within minutes of PR creation. We do not compare this against a matched human baseline, which we do not have. Second, and counter-intuitively, among pairs with complete timestamps, cross-product pairs had a lower observed median latency than same-product pairs (\num{1.2} vs.\ \num{4.7} minutes), the opposite of a ``product rubber-stamps its own'' reading. The reviewer-level results suggest that this aggregate difference is compositional: reviewers with shorter median latencies occur more frequently in the cross-product dataset, whereas Copilot, which has a longer median latency (\num{17.7} minutes, p75 \num{74} minutes), dominates the same-product dataset. Because the reviewer populations differ, the observed difference should be interpreted as reflecting reviewer composition rather than any inherent preference for same- or cross-product review.

\vspace{-10pt}
\begin{rqbox}
\textbf{RQ2 Summary:} Among PR--reviewer pairs with complete, nonnegative 
timestamps, the observed median latency was \num{1.2} minutes for cross-product 
pairs and \num{4.7} minutes for same-product pairs, although differential 
timestamp availability and reviewer composition limit this comparison. 
CodeRabbit's \texttt{refactor} share ranged from approximately 
\SIrange{10}{35}{\percent} across the five analyzed authoring agents. Three 
reviewer bots produced \SIrange{58}{65}{\percent} more comments on same-product 
PRs on average, although the distributions were long-tailed and unmatched.
\end{rqbox}
\section{Discussion}
\label{sec:discussion}

\medskip\noindent\textbf{Closed-loop AI review is a fast-growing part of the development environment.}
AI-authored and AI-reviewed pull requests, though still a minority of public GitHub activity, are already frequent enough in absolute terms that empirical software engineering can no longer assume a purely human-authored population. This complicates sampling assumptions: studies must explicitly account for AI-generated artifacts or risk mixing fundamentally different populations.

\medskip\noindent\textbf{Reviewer output differs across authoring-agent groups.}
With CodeRabbit held constant, the observed comment-category distributions differ across authoring agents. These differences may reflect the types of PRs produced by each agent, repository context, or reviewer behavior, which the present analysis cannot separate. Downstream tooling that summarizes or aggregates AI review output may therefore produce systematically different interpretations for different AI coders, even under the same reviewer.

\medskip\noindent\textbf{Same- and cross-product patterns may reflect product design.}
The two groups differ in reviewer composition and may also differ in product integration and triggering, so the observed gaps need not reflect review strictness or reviewer preference. Product boundaries also need not match model boundaries, since different products may share a foundation model and one product may run several.

\section{Threats to Validity}
\label{sec:threats}

We organize the threats to validity into construct, internal, and external dimensions.

\smallskip\noindent\textbf{Construct validity.}
Our attribution uses signature-based identification, which can undercount agents whose signatures are missing, uncatalogued, or removed by vendors. The framework is designed to prioritize precision, but residual misclassification remains possible when signatures are copied, quoted, modified, or shared across product variants. Missing or removed signatures additionally produce under-attribution. The absence of a signature is not evidence that a PR was not AI-authored or AI-reviewed: such PRs are not quarantined but simply never enter our population. Under-attribution is also asymmetric, being far lower on the reviewer side, where the major reviewers post under vendor-controlled bot logins, than on the author side, which depends on body trailers that agents may not emit and that squash merges may strip. Absolute counts are therefore lower bounds, whereas rates expressed per agent-authored PR inherit the coverage of both sides and should be read as describing the attributable population rather than all AI-authored PRs. In addition, our RQ2 content measures rely on CodeRabbit’s self-declared and observed comment categories. These labels describe the tool’s own output and are not independently validated measures of issue type, severity, correctness, or code quality. External validation with human annotation or LLM-based judges is deferred to future work.

\smallskip\noindent\textbf{Internal validity.}
Our study reports observational associations rather than causal effects. 
Confounding factors such as pull request size, language composition, 
repository-level differences, and product-specific integration behavior may 
influence the observed patterns. Timestamp completeness also differed 
substantially in the latency analysis: \SI{79.2}{\percent} of cross-product 
PR--reviewer pairs and \SI{31.9}{\percent} of same-product pairs were retained. 
If timestamp availability is associated with reviewer or workflow 
characteristics, the complete-case latency estimates may not represent the 
full same- and cross-product populations. We therefore interpret the latency 
comparison descriptively and do not attribute the observed difference to 
product pairing.

\smallskip\noindent\textbf{External validity.}
Our dataset covers public GitHub events in \dataset~(CodAGE)~\cite{codage} from 2024-01-01 to 2026-04-15, excluding private repositories, GitHub Enterprise deployments, and other version control platforms. AI reviewer activity is therefore conditioned on repositories that adopt specific integrations, and results may not generalize to non-adopting projects. Our ``closed-loop'' definition requires an AI coding agent on each side of the PR and does not require the absence of human review: the review stream we mine is restricted to AI-attributed events, so we cannot determine whether a human also reviewed a given PR, and some closed-loop PRs likely received human review as well. Finally, ``cross-product'' reflects product-level attribution rather than model-level identity.

\section{Conclusion and Future Work} \label{sec:future}

This paper presented a large-scale dataset and empirical study of AI reviews on AI-authored GitHub pull requests, defined as PRs authored by one identified AI coding agent and reviewed by one or more identified AI reviewers. Although still a minority of agent activity, such review grew rapidly over 2025 and already spans tens of thousands of cross-product interactions. Reviewer output differed across authoring-agent groups and author--reviewer 
configurations in comment-category mix and per-PR comment volume. Among pairs 
with complete timestamps, observed latency also varied across reviewer and 
product configurations, although differential timestamp availability limits 
this comparison. These findings describe observable reviewer behavior, not review correctness or software quality.

\smallskip
\noindent\textbf{Validating review content and usefulness.} Our content analysis relies on CodeRabbit's own categories, which describe its output but are not validated measures of correctness, severity, or usefulness, and prior work finds that AI review comments may be rejected, redundant, or poorly grounded in the code~\cite{lin2026coderabbit,tantithamthavorn2026hallujudge,zeng2026swrbench}. Future work should have human annotators rate a stratified sample of comments for correctness, localization, severity, and actionability, reporting inter-rater agreement, with LLM-based evaluation used only as a secondary check. Comment volume, latency, and category mix do not show whether a review improves the code, so annotation should be paired with outcome linkage to accepted suggestions, later commits, continuous-integration results, merges, and reverts, treating acceptance as a response signal rather than a measure of precision. Matched human- and AI-authored pull requests—aligned on repository, language, task type, and change size, would isolate AI authorship patterns from those of the review system or context.

\smallskip
\noindent\textbf{Separating product effects from model effects.} Attribution here operates at the product level because public records rarely reveal the underlying model, so same- and cross-product configurations are not same- and different-model configurations. Because LLM evaluators may favor outputs that resemble their own or appear more familiar~\cite{panickssery2024selfpreference,wataoka2024selfpreference,chen2025beyond}, controlled experiments could review functionally equivalent pull requests from known models under fixed configurations, using source blinding and function-preserving transformations of naming, formatting, or structure. Outcomes should include defect detection, false-positive rates, severity calibration, comment volume, and review correctness.

\smallskip
\noindent\textbf{Testing correlated failures and security risks.} AI authors and reviewers may share failure modes when they rely on related models, training sources, prompts, or tools, which the current dataset cannot assess because model identity and ground-truth defects are unavailable. A controlled benchmark could inject known functional defects and security vulnerabilities into realistic pull requests and compare same-model, related-model, and independent-model reviewers on defect recall, false positives, localization, severity, and repair correctness, showing whether reviewer diversity reduces correlated blind spots.

\smallskip
\noindent\textbf{Provenance and longitudinal measurement.} Public records reveal little about how a contribution was produced and reviewed, so future tools should expose structured metadata on authoring and reviewing products, model versions, review triggers, configurations, and human involvement, distinguishing independent review from automatic same-product review. Because the ecosystem changes rapidly, the pairings and activity levels reported here describe one period; the broader contribution is a reproducible method for measuring AI participation in authorship and review, which, repeated over time with model-level provenance and outcome-based evaluation, will help assess whether these workflows improve software correctness, security, and maintainability.

\section*{Data Availability}

The replication package, including the agent signature registry,
data-processing scripts, classifier, and analysis code, is available at
\url{https://github.com/Niruthiha/AI-AI-CodeReviews}. The source dataset,
\dataset (Coding Agent-generated GitHub Events), is available on
Hugging Face at \url{https://huggingface.co/datasets/taher-ghaleb/CodAGE} under CC BY 4.0.
These artifacts reproduce the
counts, tables, and figures reported in this paper.

\clearpage
\bibliography{refs}
\end{document}